# Strong plasmonic fluorescence enhancement of individual plant light-harvesting complexes

Farooq Kyeyune,[a] Joshua L. Botha,[a] Bertus van Heerden,[a] Pavel Malý,[b,c] Rienk van Grondelle,[b] Mmantsae Diale,[a] and Tjaart P. J. Krüger*[a]

[a] Department of Physics, University of Pretoria, Pretoria 0028, South Africa
[b] Department of Physics and Astronomy, Vrije Universiteit Amsterdam, 1081 HV Amsterdam, The Netherlands
[c] Institute of Physics, Faculty of Mathematics and Physics, Charles University, Ke Karlovu 5, 121 16 Prague, Czech Republic

**Correspondence Author** – Tjaart P. J. Krüger, tjaart.kruger@up.ac.za.

*Abstract*- Plasmonic coupling of metallic nanoparticles and adjacent pigments can dramatically increase the brightness of the pigments due to the enhanced local electric field. Here, we demonstrate that the fluorescence brightness of a single plant light-harvesting complex (LHCII) can be significantly enhanced when coupled to single gold nanorods (AuNRs). The AuNRs utilized in this study were prepared via chemical reactions, and the hybrid system was constructed using a simple and economical spin-assisted layer-by-layer technique. Enhancement of fluorescence brightness of up to 240-fold was observed, accompanied by a 109-fold decrease in the average (amplitude-weighted) fluorescence lifetime from approximately 3.5 ns down to 32 ps, corresponding to an excitation enhancement of 63-fold and emission enhancement of up to 3.8-fold. This large enhancement is due to the strong spectral overlap of the longitudinal localized surface plasmon resonance of the utilized AuNRs and the absorption or emission bands of LHCII. This study provides an inexpensive strategy to explore the fluorescence dynamics of weakly emitting photosynthetic light-harvesting complexes at the single molecule level.

*Keywords*- photosynthetic light-harvesting complexes, gold nanorods, localized surface plasmon resonance, plasmonic fluorescence enhancement, single molecule spectroscopy

## I. INTRODUCTION

Light-harvesting complex II (LHCII) is the most abundant pigment-protein complex on the earth and the main antenna complex in photosystem II (PSII) of plants and green algae[1]. It contains more than 50% of the chlorophyll (chl) molecules present in the chloroplast. The function of LHCII is twofold; firstly, under sufficiently low light intensities it absorbs sunlight and transfers the electronic excited states to the charge-separating reaction centre. Secondly, under high light intensities, it plays a photoprotective role during which excess absorbed photoenergy is dissipated in the form of heat, a process generally known as non-photochemical quenching (NPQ) of chlorophyll fluorescence. The rapidly reversible, energy-dependent component of NPQ is thought of to be regulated primarily via photophysical, energy transfer and conformational changes within the LHCII complex[2-4].

Single molecule spectroscopy (SMS) provides a unique perspective on these changes by providing access to spectroscopic information of individual complexes that would otherwise be averaged out in ensemble measurements. For example, SMS has revealed strong intensity and spectral fluctuations of various pigment-protein complexes upon continuous light illumination[5-10]. Despite the extensive studies, the mechanisms underlying NPQ have not been fully resolved and are still a topic of intense debate. One major challenge that limits the amount of information that can be retrieved from the individual quenched complexes in the context of NPQ is weak emission. Undoubtedly, an improvement in the emission brightness will help to unravel the spectroscopic properties of these complexes at the single molecule level.

Crystallography has revealed that LHCII occurs in a trimeric form, where each monomer contains eight chls *a*, six chls *b* and four carotenoids (two luteins, one neoxanthin, and one violaxanthin or zeaxanthin)[11]. The pigment confinement in LHCII is such that strong excitonic coupling amongst the pigments occurs, which leads to the high efficiency of excitation energy transfer and rapid energy equilibration within the complex[12]. Because of its high energy transfer efficiency and nanoscale dimensions, LHCII has recently been used as a building block in bio-inspired organic solar cells[13,14] and bio-photosensitizers[15,16]. However, a significant limitation of LHCII in these applications is the relatively small portion (less than 1%) of solar energy that can be absorbed by a single protein monolayer. Moreover, the intrinsic fluorescence quantum yield (QY) of LHCII in aqueous solution (~0.26)[17,18] is relatively low compared to that of the common fluorophores.

Plasmon-induced changes in the optical properties of pigments have attracted much attention in various research areas such as biosensing,[19,20] high-resolution microscopy,[21,22] photosynthesis,[23-25] and photovoltaics[26,27]. This diversity of applications is due to the ability of metallic nanoparticles (NPs) to modify the optical properties of pigments in close proximity. For example, metallic NPs can lead to large plasmonic fluorescence enhancements (PFEs) of low quantum efficiency (i.e., poorly emissive) pigments[28-32]. Significantly enhanced fluorescence upon coupling of pigments with metallic NPs arises from two factors. On the one hand, amplification of the local electric field induced by the localized surface plasmon resonance (LSPR) of metallic NPs results in significantly enhanced excitation rate of the pigments. On the other hand, the presence of metallic NPs manipulates the local density of optical states of the nearby pigments, thereby enhancing the radiative rates (the Purcell effect), which in turn leads to a change in the fluorescence lifetime and QY of the pigments[20,33]. In general, PFE depends on several factors, including nanoparticle shape, size, position and orientation of the pigment with respect to a metallic nanoparticle. Moreover, PFE strongly depends on the spectral overlap between the localized surface plasmon resonance band of the nanoparticle and the pigment absorption/emission bands[34].

In recent years, plasmonic interactions of the photosynthetic peridinin-chl-*a* protein (PCP) and purple bacterial light-harvesting complex 2 (LH2) with metallic nanostructures such as silver island films, gold and silver nanospheres, and gold nanorods have been explored[25,35-37]. Van Hulst and co-workers have demonstrated PFE of up to 523-fold, accompanied with a 10-fold increment in photostability of single LH2 complexes randomly coupled to lithographically patterned gold nanoantennas[25]. In another approach by Kaminska et al., single PCP monomers were coupled to self-assembled DNA origami-based gold spherical dimers of 100 nm diameter, and PFEs of slightly over 500-fold were similarly reported[35]. These works have demonstrated the ability of metallic NPs in drastically manipulating the optical properties of single photosynthetic light-harvesting complexes. However, the utilized methods involved sophisticated and expensive fabrication processes. There is a need to develop simple, scalable, inexpensive yet effective methods to construct similar biological-metallic hybrid antenna systems. Moreover, plasmonic interactions with plant complexes have so far not been reported.

Herein, we present a study on the fluorescence dynamics of individual LHCII complexes coupled to gold nanorods (AuNRs) using SMS at room temperature. The samples were excited at a wavelength of 646 nm within the longitudinal plasmon band of AuNRs but below the fluorescence band of LHCII. For each measurement, we simultaneously recorded the emission spectrum, brightness and fluorescence lifetime. This approach allowed us to correlate changes in all three measured variables. Notably, the results reveal a maximum of ~240-fold PFE in the brightness of individual LHCII complexes accompanied by lifetime shortening (down to ~32 ps, at the temporal resolution limit of the experimental setup).

## II. RESULTS AND DISCUSSION

Fig. 1a shows the statistical distribution of the aspect ratio of about 100 AuNRs, measured using a transmission electron microscope (TEM). The nanorods featured a narrow size distribution with an average aspect ratio (length divided by width) of 2.3 ± 0.2. The absorption and emission spectra of LHCII in buffer solution are compared to the absorption of AuNRs in Fig. 1b. The LHCII absorption bands peaking at 436 and 675 nm originates from Chl *a* and those peaking at 473 and 650 nm mainly

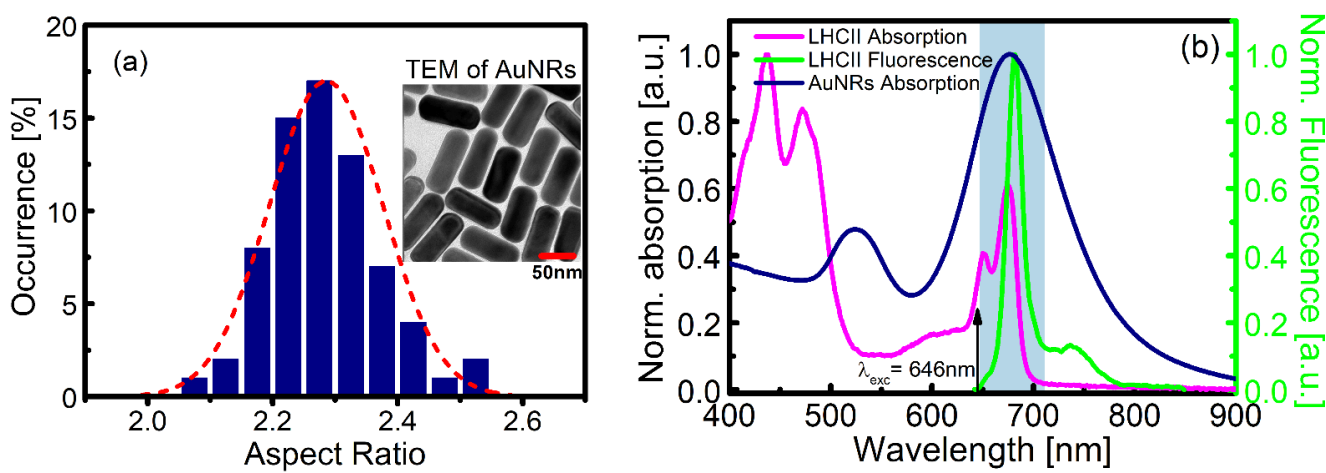

Fig. 1 (a) Aspect ratio distribution of the AuNRs, with a Gaussian fit (ref, dashed) peaking at 2.3 ± 0.2 nm. The inset shows a transmission electron micrograph of the AuNRs before dilution, with the mean length and width being 85 ± 4 nm and 37 ± 4 nm, respectively. (b) Normalized ensemble absorption (magenta) and fluorescence (green) spectra of LHCII, as well as absorption (blue) spectrum of the colloidal AuNRs in 25 mM PSS. Notice the spectral overlap (light blue region) between the longitudinal localized surface plasmon band of the AuNRs and the absorption and emission of LHCII complexes.

from Chl *b*. The emission spectrum has a typical maximum at 680 nm. The nanorods exhibit two plasmon bands, i.e., a transverse localized surface plasmon resonance band at around 525 nm and a longitudinal localized surface plasmon resonance (LLSPR) band peaking at 673 nm. The aspect ratio of the AuNRs was chosen such that their absorption spectra overlap optimally with the absorption and emission bands of LHCII complexes. This was achieved by adjusting the concentration of silver ions during the synthesis process. It should be noted that strong spectral overlap facilitates efficient coupling of the plasmons in the AuNRs with the photosynthetic complexes in the hybrid system.

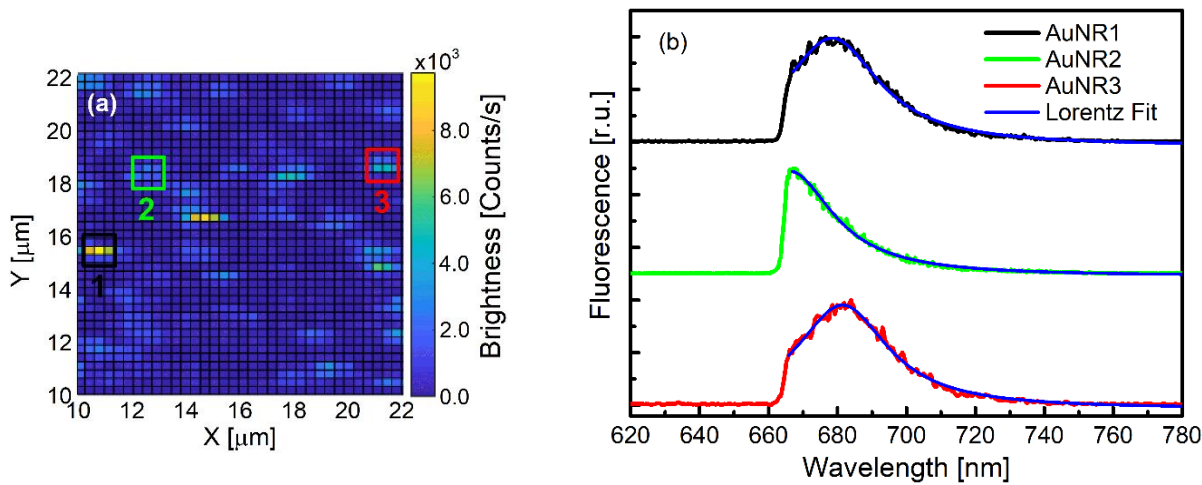

Fig. 2 (a) Typical 12 μm × 12 μm fluorescence image of individual AuNRs immobilized on functionalized glass substrate. The excitation intensity was 258 W/cm$^2$ at 646 nm (b) Representative fluorescence emission spectra of three randomly selected AuNRs labelled in (a). Spectra are normalized and offset for clarity. Blue lines show Lorentzian fits. The spectra were fitted up to the sharp cut-off of the fluorescence filter.

For control purposes, the emission properties of immobilized AuNRs were characterized first. The AuNRs were excited at 646 nm with an excitation intensity of 258 W/cm$^2$. A typical 144 μm$^2$ fluorescence image of the isolated AuNRs on a glass substrate is shown in Fig. 2a. The density of the nanorods was 15 - 25 AuNRs/144 μm$^2$. Fig. 2b shows the emission spectra of three selected individual AuNRs. The spectral peak position of the individual AuNR s changes from one nanorod to the other. This heterogeneity in the peak positions can be attributed to two factors: the distributions in the aspect ratio (Fig. 1a) and the deviations in tip curvature of individual AuNRs. Both factors lead to a distribution in the absorption cross section of individual AuNRs. The emission spectrum of AuNRs excited near the LLSPR invariably follows the surface plasmon band of the nanoparticle. Consequently, AuNRs usually exhibit Stokes emission accompanied by anti-Stokes emission[38,39]. This is evident from the spectrum of AuNR2 in Fig. 2b. Fitting of about 38 individually measured Stokes-shifted spectra with a Lorentzian function reveals, on average, a spectral width (full-width at half-maximum, FWHM) of 29 ± 2 nm.

Next, the optical properties of individual LHCII and LHCII coupled to AuNRs (LHCII@AuNRs) were examined, after excitation at 646 nm with intensities of 70.8 W/cm$^2$ and 17.7 W/cm$^2$, respectively. The excitation intensity of LHCII@AuNRs was reduced to slow down photobleaching of LHCII that would otherwise occur rapidly due to the strong near-field produced by the nanorods. The concentration of LHCII in both samples was determined empirically to achieve on average, 7 - 10 complexes per 100 μm$^2$. Fig. 3a depicts an exemplar fluorescence brightness image of isolated, immobilized LHCII complexes. The image features randomly distributed spots of similar size. For each spot, a brightness-time trace was measured and a representative trace is displayed in Fig. 3c. The brightness counts were

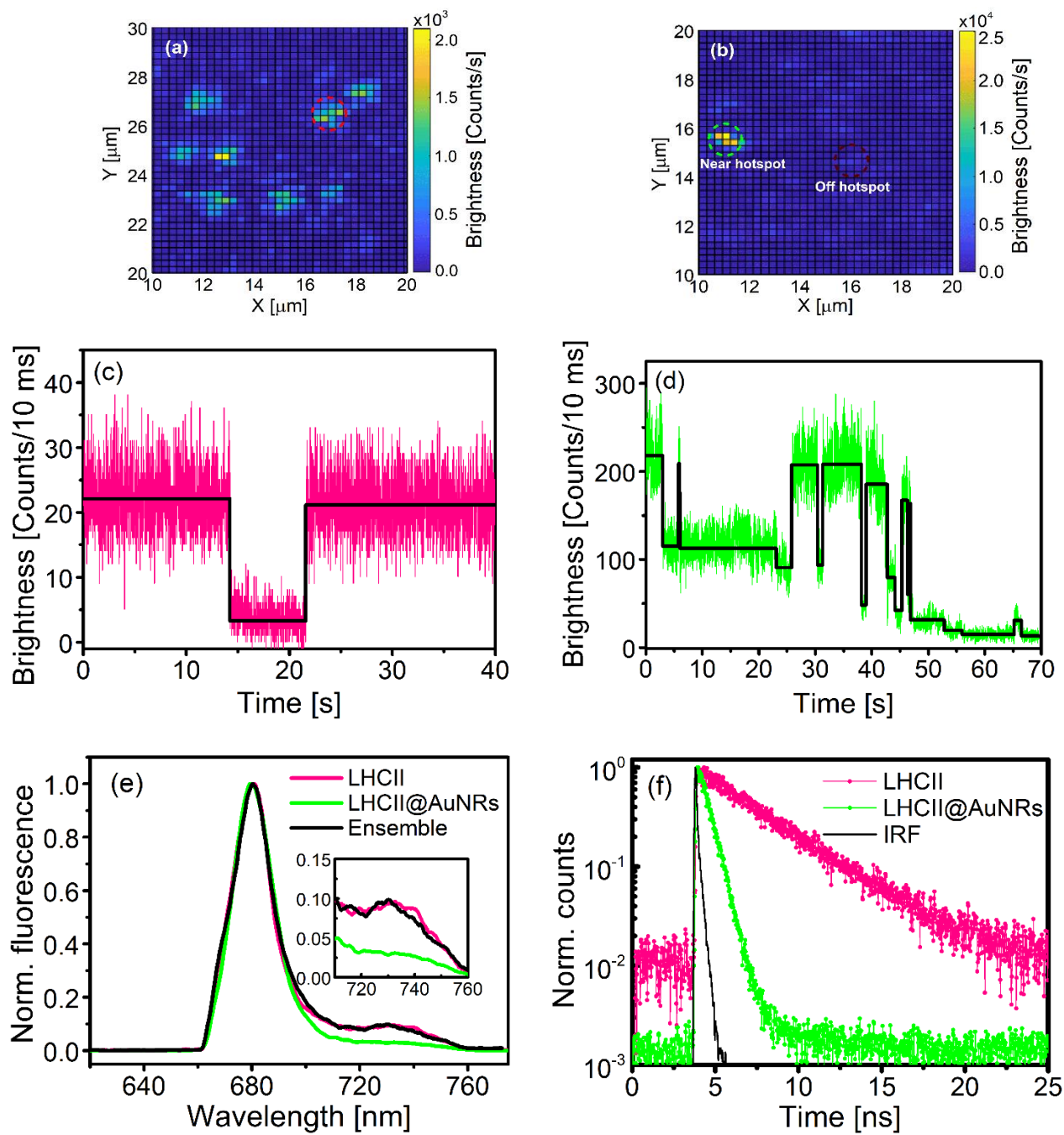

Fig. 3 (a, b) Typical 10 μm × 10 μm raster-scanned brightness images of single LHCII and LHCII@AuNRs complexes. The excitation was at 646 nm with intensities of 70.8 W/cm$^2$ (a) and 17.7 W/cm$^2$ (b), respectively. The very bright spot in the green circle depicts a single LHCII complex near a plasmonic "hot spot" and the spot in the dark brown circle shows unenhanced LHCII complex. (c, d) Fluorescence time traces of single LHCII and LHCII@AuNRs. (e) Examples of one-second integrated emission spectra of LHCII (sum of 40 individual spectra, pink), LHCII@AuNRs (sum of 38 individual spectra, green) and ensemble emission of LHCII (black). The inset shows a magnification of the vibrational bands. (f) Fluorescence decay traces of encircled complexes in (a) and (b).

integrated into bins of 10 ms and resolved using an intensity change-point algorithm[40]. A complex near the hot spot of a nanorod exhibits a bright fluorescence spot due to its interaction with the dipolar plasmon mode, as shown in Fig. 3b by the spot encircled in green. The corresponding fluorescence transient in Fig. 3d indicates a PFE of about 12-fold compared to one of the unenhanced complexes (brown encircled dim spot in Fig. 3b). Additional traces of single LHCII and LHCII@AuNRs complexes are shown in Fig. S1. All measured complexes showed blinking or single-step photobleaching dynamics that occurred considerably faster than the 10-ms bin times in Figs. 2b-c and S1. This behavior gave evidence that the observed intensity profiles originated from single complexes with well-connected pigments[30,41]. To observe the spectral signature as a result of the plasmonic coupling of LHCII complexes near AuNRs, we recorded a series of consecutive emission spectra with an integration time of one second each. Fig. 3e shows a comparison of the sum of the measured spectra for the complexes highlighted in Fig. 3b. The emission spectra of both samples match the bulk steady-state emission spectrum of LHCII, with the emission band centered at 680.5 nm. The effect of plasmonic coupling is evident in the fluorescence decay trace shown in Fig. 3f, where the fluorescence lifetime of LHCII@AuNR was significantly shortened by nine times (3.45 ns for the unenhanced complex versus 360 ps for the enhanced complex).

The fluorescence brightness of individual pigments is dependent on the excitation intensity and the intrinsic fluorescence QY[42]. In Fig. 4a we varied the excitation intensity of individual LHCII complexes and recorded the corresponding fluorescence brightness. Each data point was obtained by taking the average of the ON state (Fig. 3c) brightness levels of about 20 ± 5 complexes. The complexes were measured over an excitation period of 20 seconds under continuous pulsed laser irradiation. The data follow a general three-level model[18,43] $I_\infty I_e/(I_s+I_e)$, where $I_s$ is the saturation intensity, $I_e$ is the excitation intensity and $I_\infty$ is the maximum achievable fluorescence brightness during the ON state. In LHCII, intersystem crossing has a yield of about 30 – 40%[18,44]. An excitation in the chl triplet state is rapidly transferred to the triplet state of a nearby carotenoid that has a relatively long lifetime of several microseconds[45]. Under intense illumination, the resulting excited triplet carotenoid plays a photoprotective role by quenching the chl singlet excited states via singlet-triplet annihilation[46]. This mechanism is excitation intensity dependent, hence limiting the maximum attainable fluorescence brightness at high excitation photon density (Fig. 4a). Plasmonic coupling can reduce the amount of singlet-triplet annihilation in the following ways: shortening of the fluorescence lifetime, so that the excited triplet yield is reduced, enhancement of the fluorescence brightness at reduced excitation intensity, and a reduction in the excitation volume (i.e., forming so-called "hot spots") to increase the overall emission collection efficiency.

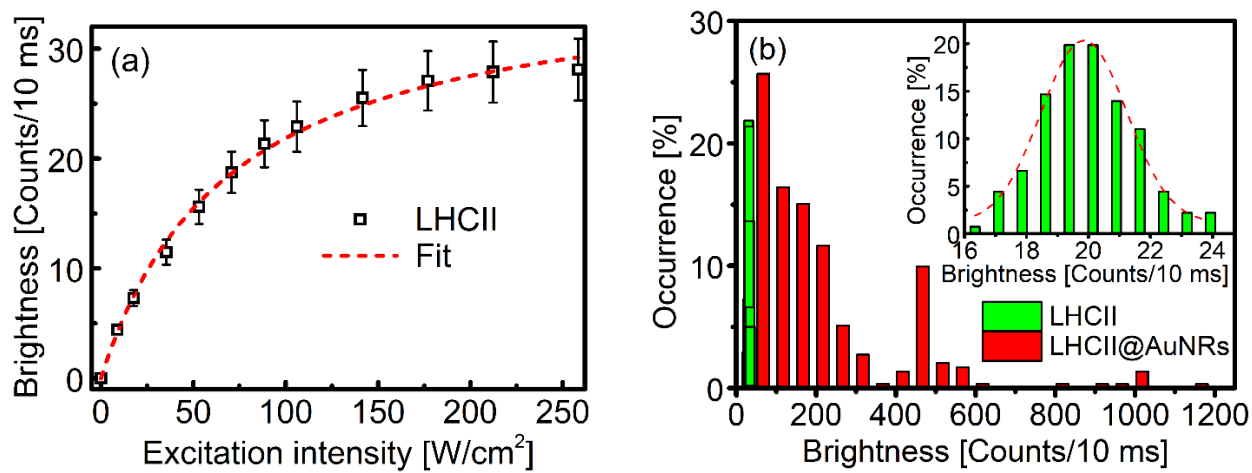

Fig. 4 (a) Saturation curve of the brightness of individual isolated LHCII complexes. For each excitation intensity, the number of measured complexes was 15 – 25. The black open squares are the mean values and the error bars illustrate the standard deviation. The data was fitted with a three-level model with the fitting parameters $I_s = 74 \pm 2.8$ W/cm$^2$ and $I_\infty = 36.8 \pm 5.5$ counts/10 ms. (b) The statistical distribution of the fluorescence brightness of LHCII and LHCII@AuNRs. The green bars represent the brightness distribution of individual LHCII complexes excited at 70.8 W/cm$^2$, while the red bars represent the brightness distribution of LHCII@AuNRs excited at 17.7 W/cm$^2$. The inset shows the magnified fluorescence distribution of LHCII. Bins of 1 count/10 ms and 50 counts/10 ms were used for LHCII and LHCII@AuNRs, respectively.

To assess the PFE of individual LHCII, the fluorescence brightness corresponding to the unquenched (or ON) state of the complexes in the two environments was compared, as shown in Fig. 4b. The brightness of 130 individually measured unquenched LHCII complexes followed a Gaussian distribution with a mean value of 19.8 ± 0.1 counts/10 ms. Upon coupling of individual LHCII complexes to nanorods, their brightness was strongly enhanced. A significant variation in the relative brightness of LHCII@AuNRs of up to 1200 counts/10 ms can be observed. This variation is attributed to different aspect ratios of the AuNRs and differing spectral overlap and coupling distance between the AuNRs and isolated LHCII complexes as well as the orientation of LHCII within the region of hot spot. Chemical synthesis of AuNRs results in a distribution in the aspect ratio of the AuNRs (Fig. 1a) and a corresponding variation in plasmonic interactions. For all the hybrid nanostructures, the minimum total distance between a single LHCII and AuNR was roughly 4.8 nm. This is the sum of the CTAB bilayer on the AuNR surface (2.9 nm) and PSS layer (1.9 nm)[47] used to immobilize the AuNRs. Additionally, since the interaction between pigments and metallic nanoparticles is strongly distance dependent[48], we can intuitively attribute the observed effects to the variation in coupling distance between the individual LHCII complexes and the hot spots near the tips of the AuNRs.

The results shown in Fig. 4b were used to calculate the PFE factors due to the plasmonic coupling in the LHCII@AuNRs hybrid system, using $PFE = I'I^0_{exc}/I^0_{avg}I'_{exc}$, where $I'$ is the measured brightness of LHCII@AuNRs, $I^0_{exc}$ is the excitation intensity of LHCII, $I^0_{avg}$ is the average brightness (19.9 ± 0.1 counts/10 ms) of LHCII, and $I'_{exc}$ is the excitation intensity of LHCII@AuNRs. We note that this equation can give artificially increased PFE if the reference sample is excited in the saturation regime were the maximum achievable fluorescence is reduced, see Fig. 4a. The calculated PFE factors are presented in Fig. 5b–c, with a maximum of 242. Higher PFE of individual LH2 (~500) and PCP (~526) complexes coupled to lithographically patterned gold nanorods and DNA origami-based metallic nano-antennas, respectively, were previously reported[25,35]. It should be noted that the fluorescence QY of LH2 and PCP complexes are ~0.10 and ~0.11, respectively, which are <50% than that of LHCII, which justifies the lower PFE of LHCII as compared to those of LH2 and PCP. However, if we adopt the figure-of-merit used in ref 32 (for unbiased comparison), FOM = PFE × intrinsic QY, then our results are in good agreement with the previous reports. The large enhancement factor reported here is due to the high degree of spectral overlap between the LLSPR band of the AuNRs with the emission and

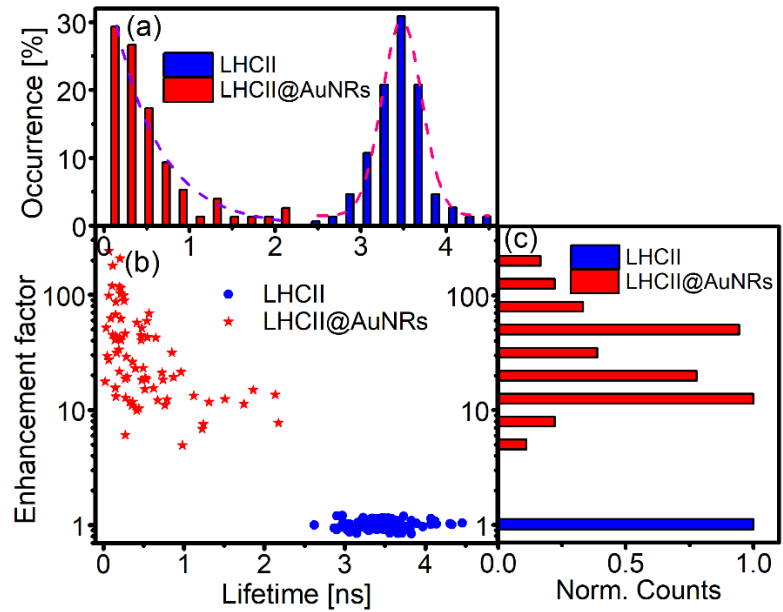

Fig. 5 (a) Statistical distribution of the average fluorescence lifetimes of about 80 LHCII@AuNRs (red bars) and 130 LHCII (blue bars), with bin size of 0.2 ns. The dashed lines show a fitted Gaussian (pink) and mono-exponential decay function (violet) for LHCII and LHCII@AuNRs, respectively. (b) Scatter plot of PFE factor versus fluorescence lifetime of LHCII@AuNRs (red stars) and isolated LHCII complexes (blue circles). (c) Distribution of the PFE factor of LHCII@AuNRs (red bars) and mean value of the brightness of LHCII complexes (blue bars).

absorption band of LHCII complexes (Fig. 1b). The advantage of our approach is that it does not require complicated and expensive fabrication processes to construct LHCII@AuNR nanostructures.

Enhancement of the excitation field and expediting of the radiative decay rate can both lead to the observed fluorescence enhancement. To gain more insight into the corresponding enhancement rates, we measured the fluorescence lifetimes of single coupled LHCII@AuNRs. For this purpose, intensity decay traces of several uncoupled and coupled LHCII complexes were recorded in time-tagged time-resolved (TTTR) mode. The fluorescence lifetimes were determined by fitting the decay traces with monoexponential (for LHCII), or triple-exponential (for LHCII@AuNRs) functions The latter was needed to describe the fractional contribution of decay times of different components arising from the pigment-surface plasmon interactions. For only a small fraction (< 1%) of LHCII complexes, a double exponential function was required to improve the fitting, in which case the second component was attributed to the contribution from another complex inside the excitation focal volume. An acceptable $\chi^2$ was used to test the goodness-of-fit. For multi-exponential fitting, the average lifetime was calculated using the amplitude weighting for each decay trace. Fig. 5a shows the fluorescence lifetime distribution of LHCII and LHCII@AuNRs. The mean lifetime of about 130 unquenched LHCII complexes was found to be 3.5 ± 0.3 ns, which is in good agreement with the reported fluorescence lifetime of single LHCII in solution[5] or adsorbed on poly-L-lysine-coated glass substrates[44]. This result shows that LHCII retains its functionality as an efficient light-harvester.

In contrast, the LHCII@AuNRs hybrid sample presented remarkably shorter amplitude-averaged lifetimes down to ≤ 32 ps (Fig. 5a–b), accompanied by strongly enhanced fluorescence brightness, a signature of plasmonic interaction of pigments with the metallic NPs, as indicated in the Supplementary Equations. This observation can be explained by the fact that the strongest PFE is obtained at the shortest distance of the LHCII complex from the nearby AuNR near one of its tips where both the excitation and radiative rates are strongly enhanced. As the LHCII complex comes close to the AuNR, part of its emission is transferred to the nanoparticle through plasmonic resonance energy transfer, which results in quenching[49], a feature accompanied by a reduction in both fluorescence lifetime and brightness. However, our selection criterion excluded from the subsequent analysis all complexes whose brightness was below the threshold (7.2 counts/10 ms). The large PFE factors indicate that most of the energy remains in the antenna complexes without being transferred to the nanorods or thermally dissipated via a strong increase in the non-radiative rate.

To clarify the individual contributions to the PFE, a semi-empirical model[50] was used to estimate the excitation and emission parts of the enhanced brightness of LHCII near the hot spot of a AuNR. The modified fluorescence QY in the presence of a metallic NP is given by $Q_m = \gamma_m/(\gamma_m + \gamma_{nr,m})$ (see the supporting equations for details). Then, the fluorescence emission enhancement can be estimated from $E_{em} = Q_m/Q_0$, and the rest of the overall PFE is attributed to the excitation enhancement: $E_{exc} = PFE/E_{em}$. These factors were calculated based on the intrinsic QY of LHCII (0.26) in solution and the measured fluorescence lifetime of LHCII with or without AuNRs. Since the minimum space between the AuNR surface and a nearby LHCII was estimated to be 4.8 nm, we assumed that this distance is enough to prevent fluorescence quenching due to ohmic losses into the metal. Thus, the non-radiative decay rate is constant and does not change from complex to complex. The calculated excitation and emission enhancement factors show that the maximum PFE was achieved when the excitation rate was enhanced by 63-fold with an emission efficiency that corresponds to an increase by 3.8-fold (see Fig. S2a). In addition, the radiative rate is enhanced by 200-fold (Fig. S2b), which corresponds to a decay rate of 15 ns$^{-1}$ versus 0.074 ns$^{-1}$ in the absence of metallic NPs.

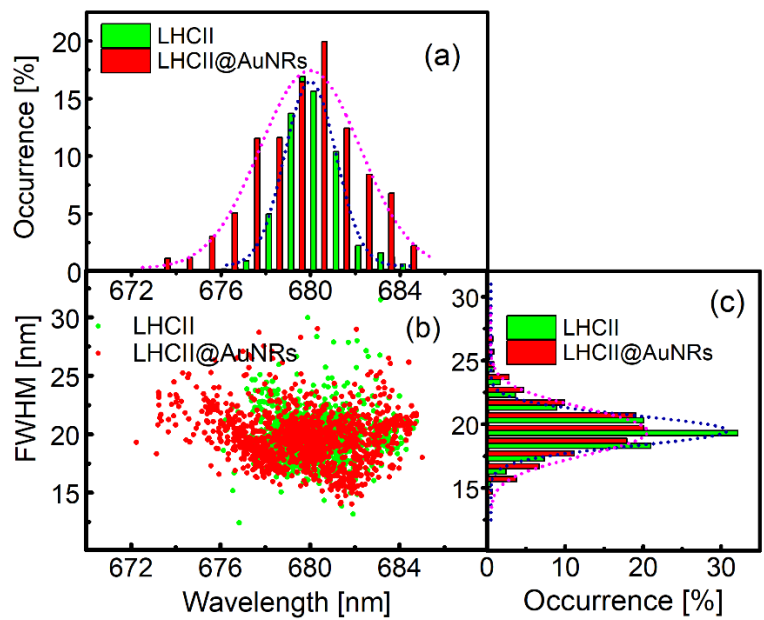

Fig. 6 (a) Fluorescence peak position distribution obtained by Gaussian fitting of the spectra of 130 LHCII (green) and 81 LHCII@AuNRs (red) measured complexes. Data was distributed in bins of 1 nm. The peak position of the Gaussian fits for both samples (dashed lines) is at 680 nm. (b) Scatter plot of peak positions versus FWHM for both samples. (c) Corresponding 1-nm binned FWHM distributions.

To gain more insight into the plasmonic coupling of LHCII with the nanorods, emission spectra of isolated individual LHCII and LHCII@AuNRs complexes were measured, and the spectral peak distributions are shown in Fig. 6a. Both distributions are centered around 680 nm, qualitatively consistent with the bulk fluorescence emission (Fig. 1b). The variation in the peak position of individual

LHCII complexes can be explained by static disorder, which changes the exciton composition in LHCII under steady-state conditions[50]. The broadening of the spectral distribution of LHCII@AuNRs is attributed to the plasmonic coupling of the nearby nanorod with the lowest exciton states of LHCII. The average spectral width (FWHM) of the complexes in both environments was 19.7 nm (Fig. 6c), consistent with previous studies[50,51]. These observations suggest that the vast majority of the measured complexes remained intact.

### III. CONCLUSION

To summarize, we have studied the effects of plasmonic coupling of individual LHCII complexes with single gold nanorods. The hybrid nanostructures were constructed using a spin-assisted layer-by-layer technique from colloidal solutions. We observed large fluorescence brightness enhancement of LHCII of up to 240-fold, which is very high considering the complex's QY of 0.26. The increase in fluorescence brightness is accompanied by >100 times fluorescence lifetime (amplitude averaged) shortening down to 32 ps. The large enhancement factors are attributed to the strong spectral overlap of the longitudinal localized surface plasmon resonance of the synthesized nanorods with the absorption and emission peaks of LHCII complexes. This work explores the possibility of designing simple, inexpensive and efficient nano-bio hybrid systems, which could have a significant impact on single molecule spectroscopy of poorly emitting pigments as well as bio-nano solar cells, quantum optics, bio-sensing, and materials science.

### IV. MATERIALS AND METHODS

**Materials:** Tetrachloroauric (III) acid ($HAuCl_4 \cdot 3H_2O$, 99.9%), cetyltrimethylammonium bromide (CTAB, 99%), sodium borohydride ($NaBH_4$, 99.99%), silver nitrate ($AgNO_3$, 99.99%), L-ascorbic acid, hydrochloric acid (37%), poly(diallyldimethylammonium chloride) (PDADMAC, MW ~200,000-350,000, 20 wt. % in $H_2O$), polystyrene sodium sulfonate (PSS, MW ~70,000), poly(vinyl alcohol) (PVA, MW ~124,000), 4-(2-Hydroxyethyl) piperazine-1-ethanesulfonic acid sodium salt (HEPES, 99.5%), n-dodecyl β-D maltoside (β-DM, MW ~50,000) were purchased from Sigma-Aldrich. All chemicals were used as received without any further purification. Milli-Q (18 MΩ.cm at 27 °C) water was used in the preparation of all solutions.

**Synthesis of AuNRs**: AuNRs were synthesized by seed-mediated protocol, as described elsewhere[52,53], with a few modifications. First, a seed solution was prepared by mixing 250 μL of 10 mM $HAuCl_4$ with 9.75 mL of 100 mM CTAB followed by vigorous stirring. To the stirred solution, 600 µL of 10 mM ice-cold (~4 °C) NaBH$_4$ was added. The solution was gently stirred until it turned yellow-brown and then left unperturbed in a water bath at 28 °C for at least two hours. Next, the growth solution was prepared by mixing 2 mL of 10 mM HAuCl$_4$ with 40 mL of 100 mM CTAB. Then, 280 µL of 10 mM AgNO$_3$ was added, followed by 720 µL of 1 M HCl. The solution was gently mixed by swelling before adding 320 µL of freshly prepared 100 mM ascorbic acid. Finally, 12 µL of as-prepared seed solution was added and gently mixed for 15 s. The mixture was then aged for 18 h in a water bath at a temperature of 28 °C. The as-prepared AuNRs were purified by centrifuging twice at 8000 rpm for 20 min each time. The obtained AuNRs were diluted in deionized (DI) water to the required concentration.

**Preparation of LHCII@AuNRs hybrids**: The LHCII trimers were isolated from spinach thylakoid membranes as previously described in the literature[54], with a few modifications to increase sample integrity. LHCII was diluted in 25 mM HEPES, pH 7.5, 0.03 % β-DM, 0.75 mg/mL glucose oxidase, 7.5 mg/mL glucose, 0.1 mg/mL catalase and 0.25 % PVA to a final concentration of 10 pM. The samples were prepared by employing a spin-assisted layer-by-layer technique using a spin coater (Model WS-650MZ-23NPPB, Laurell Technologies). First, the glass substrates were cleaned using an ultrasonic bath in acetone, ethanol and deionized water sequentially for 15 min at each step. The substrates were then dried under a stream of nitrogen gas and treated with UV-ozone for 20 min. Next, 25 mM of positively charged PDADMAC in 10 mM NaCl was spin coated onto cleaned glass substrates (200 µL, 4000 rpm, 20 s), which were subsequently rinsed copiously with Milli-Q water and dried under a stream of nitrogen gas. A mixture of CTAB-coated AuNRs and negatively charged PSS (60 µl of AuNRs and 140 µL of 25 mM PSS in 10 mM NaCl) was spin-coated (4000 rpm, 20 s) onto PDAMAC functionalized glass substrates. The substrates with the AuNRs were then rinsed with deionized water to remove unbound CTAB and again dried under a stream of nitrogen gas. This procedure gave a sparse distribution of about 15 – 25 isolated single AuNRs per 144 µm$^2$ area. Next, LHCII diluted in PVA was spin-coated (30 µL, 2500 rpm, 30 s) at room temperature (23 ± 1 °C) onto the AuNRs substrates. The LHCII complexes were randomly distributed across the PVA thin-film of thickness 15 ± 5 nm, as determined by a profilometer (Alpha-Step, Tencor Instruments). In addition, a reference sample was prepared similarly by spin coating LHCII diluted in PVA onto barely cleaned glass substrates. Experiments were carried out immediately after sample preparation.

**Spectroscopic and microscopic measurements:** Absorption spectra of LHCII and AuNRs were measured using a Cary UV-vis spectrophotometer. The AuNR size and aspect ratio were determined using a Zeiss transmission electron microscope (TEM) operating at an acceleration voltage of 100 kV. TEM samples were prepared by depositing 3 μL of AuNRs solution onto copper grids with formvar film support. TEM images of about 100 AuNRs were collected from which the aspect ratio distribution of the AuNRs was determined.

**Experimental set-up**: The particles were excited by a pulsed laser (Fianium, SC400–4–PP) with a pulse repetition rate of 40 MHz and a central wavelength of 646 nm selected by an acousto-optic tunable filter (AOTF, Crystal Technology, Inc.). The excitation light was circularly polarized by a combination of a linear polarizer (LPVISB050-MP, Thorlabs) and a quarter wave plate (λ/4 485-630, Achromatic Retarder, Edmund Optics). After passing through a spatial filter, the laser beam was directed by a dichroic mirror (TX660, Chroma) towards an oil immersion objective (1.45 NA, Plan-Fluor Apo λ 100X, Nikon), focusing it tightly to a diffraction-limited spot (FWHM of approximately 0.6 μm) on the sample mounted on a motorized piezo stage (Mad City Labs, LPS200). The fluorescence emitted by the sample was collected by the same objective and transmitted through the dichroic mirror, a 100 μm pinhole to reject out-of-focus background light, a fluorescence filter (ET665lp, Chroma) and a 30/70 beam splitter. The fluorescence spectrum (30% of the fluorescence emission) was measured using an electron multiplying charge-coupled device camera (EMCCD, Andor iXon$_3$) with an integration time of one second. The brightness (70% of the fluorescence emission) and decay traces were measured using a Micro Photon Devices PDM series single–photon avalanche photodiode (PD–050–CTE, FWHM of approximately 128 ps), and a Becker & Hickl time-correlated single-photon counting (TCSPC) module. Data acquisition was performed using a custom-written LabVIEW (National Instruments) script. A 10 μm × 10 μm area was scanned, and the positions of single isolated particles were identified. The piezo stage was moved to position the laser beam on each of these individual particles one at a time, to record the fluorescence emission. Each complex in the reference sample was measured continuously for 40 s and LHCII@AuNRs complexes were measured until photobleached.

**ACKNOWLEDGMENT**

We thank Erica Belgio for providing the LHCII sample. We also acknowledge funding from the National Research Foundation (NRF), South Africa grant no. 102431 (F.K.), grants no. 8990, 94107, 112085 and 109302 (T.P.J.K.). T.P.J.K. was further supported by the Photonics Initiative of South Africa, Rental Pool Programme of the National Laser Centre, South African, Department of Science and Technology and the University of Pretoria through the Research Development Programme, Strategic Research Funding and the Institutional Research Theme on Energy. M.D. acknowledges funding from the NRF Nanotechnology Flagship Program grant no. 88021. J.L.B. was supported by the VU University Amsterdam−NRF South Africa Desmond Tutu Programme. B.V. was supported by the Department of Science and Technology−NRF grant no. 115463. R.v.G. gratefully acknowledges his "Academy Professor" grant from the Netherlands Royal Academy of Arts and Sciences (KNAW).

## Supplementary Information

**Data analysis:**

Only brightness transients showing single step blinking and photobleaching dynamics were considered. Analysis of intensity brightness was performed using a home-written MATLAB (MathWorks) script based on an intensity change-point algorithm that uses the time-tagged photon data as input[1]. The fluorescence brightness of each complex was obtained from the first resolved intensity level, which was in all cases an ON state. For PFE, a minimum threshold for enhanced brightness was established. For this purpose, the intensity traces of dim complexes in the LHCII@AuNRs sample and their corresponding lifetimes were measured. Unenhanced complexes exhibited average brightness levels of 7.5 counts/10 ms, and typical lifetimes of 3.5 ns. The fluorescence lifetimes were obtained by fitting the decay traces with either a single exponential function (LHCII) or a multi-exponential function (LHCII@AuNRs), convoluted with the instrument response function (at $\lambda_{em}$ = 680 nm) using a home-written Python algorithm. It uses the least-square minimization strategy to find the best-fit parameters for the given data[2]. The fluorescence peak position and FWHM distributions were obtained by fitting a skewed Gaussian to all the single-molecule fluorescence spectra as previously reported[3].

**Supplementary Equations: Quantum yield and fluorescence lifetime in the presence of metallic NPs**

The plasmonic excitation and emission enhancements were analyzed using a semi-empirical model proposed in ref 4. In the absence of metallic NPs or any other quenching interactions, the intrinsic quantum yield of an isolated pigment is given by:

$$Q_0 = \frac{\gamma_r}{\gamma_r + \gamma_{nr}},$$

where $\gamma_r$ and $\gamma_{nr}$ are the intrinsic radiative and non-radiative decay rates, respectively. The fluorescence lifetime is given by the inverse of the total decay rate:

$$\tau_0 = \frac{1}{\gamma_r + \gamma_{nr}}.$$

The coupling of pigments to metallic NPs modulates both the radiative and non-radiative decay rates. The modulated quantum yield can be expressed as follows[5]

$$Q_m = \frac{\gamma_m}{\gamma_m + \gamma_{nr,m}},$$

where $\gamma_m$ and $\gamma_{nr,m}$ are the modified radiative and non-radiative rates in the proximity of metallic NPs. $\gamma_{nr,m}$ includes ohmic losses into the metal[6]. The modification of the total decay rates leads to shortening of the fluorescence lifetime, which is:

$$\tau_m = \frac{1}{\gamma_m + \gamma_{nr,m}}.$$

For simplicity, we can assume that $\gamma_{nr,m} \approx \gamma_{nr}$ when the pigment lies within an appropriate distance from the metal surface to prevent quenching of the excited state fluorescence to the metal. In our work, the minimum spacer between the AuNR surface and LHCII was estimated to be ~4.8 nm, meaning that ohmic losses into the metal were minimized. Using the measured fluorescence brightness and lifetime of LHCII@AuNRs and LHCII, intrinsic fluorescence quantum yield of 0.26 of LHCII in solution and assuming that the modified non-radiative rate in the presence of AuNRs is constant and equivalent to the intrinsic non-radiative rate, we can estimate the emission enhancment uisng the equation $E_{em} = \frac{Q_m}{Q_0}$. Then, the excitation enhancement is determined from the overal plasmonic flourescence enhancement as follows: $E_{exc} = \frac{PFE}{E_{em}}$.

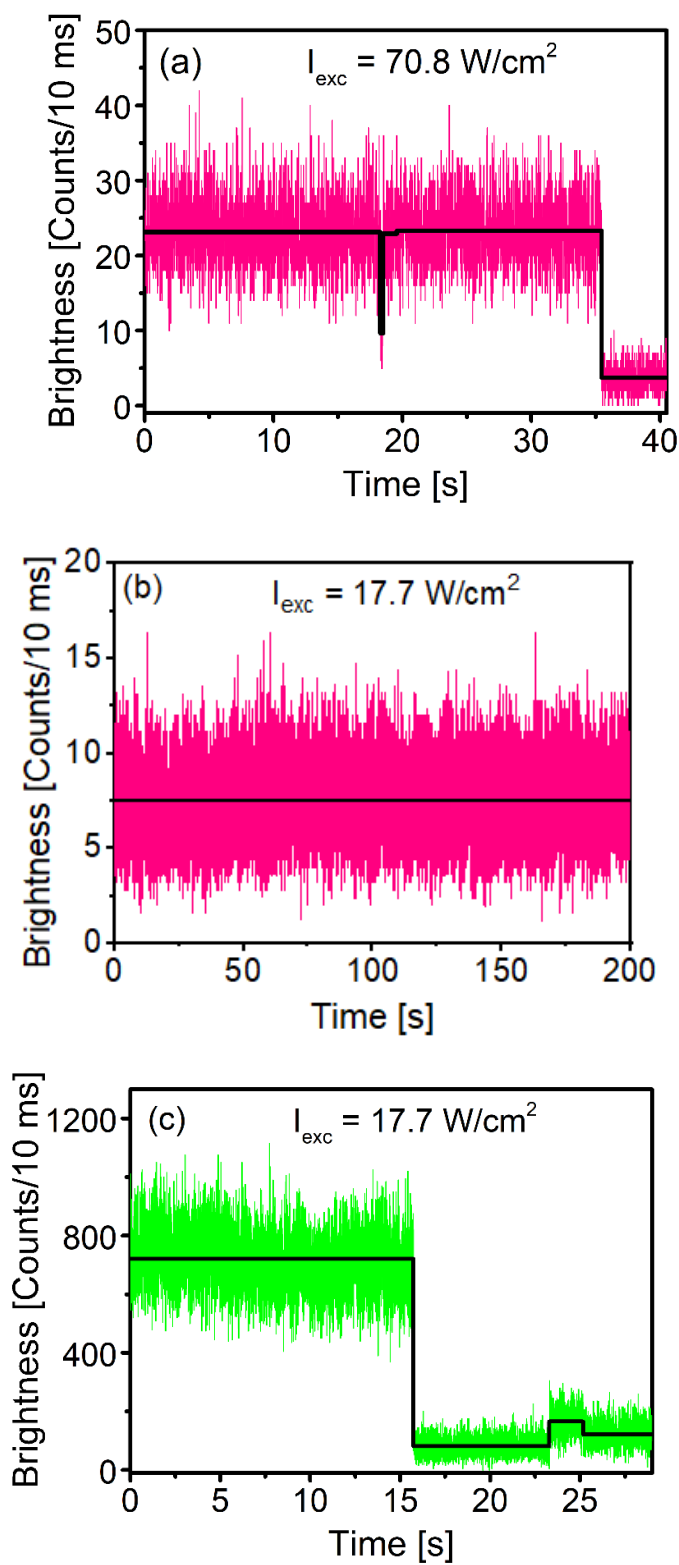

Fig. S1 Fluorescence brightness time traces of single LHCII complexes excited at 646 nm without AuNRs (a-b), and in the presence of a AuNR (c). The intensity levels (black) were obtained by a change-point algorithm after binning the time-tagged photons into consecutive 10 ms bins.

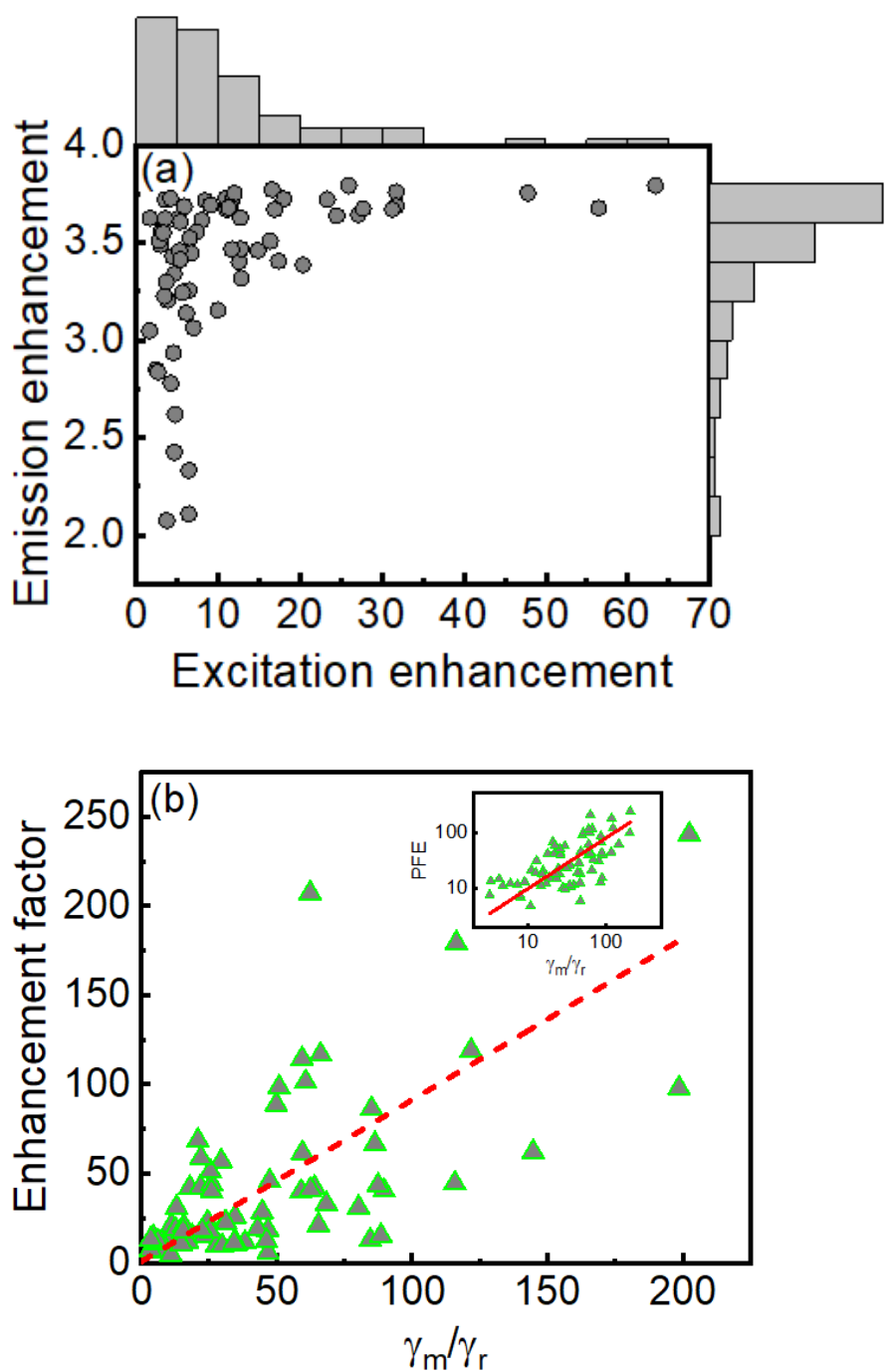

Fig. S2 (a) Correlation between excitation enhancement and emission enhancement factors. Data was distributed into bins of 5 and 0.2 for x and y-axes, respectively. (b) Correlation of plasmonic fluorescence enhancement factor and the ratio of the modified radiative rate to the intrinsic radiative rate. Insert is the same graph on a logarithmic scale, fitted with a linear function.